\documentclass[final,3p,times]{elsarticle}

\usepackage{amssymb}
\usepackage{amsmath}
\usepackage{slashed}
\usepackage{hyperref}
\usepackage{url}

\journal{Journal of Subatomic Particles and Cosmology}

\begin{document}

\begin{frontmatter}



\title{Heavy quark thermalization using phonon-heavy quark EFT}

\author[aaa]{Andreas Kirchner}
\affiliation[aaa]{organization={Physics Department, Duke University},
             addressline={Science Drive},
             city={Durham},
             postcode={27708},
             state={NC},
             country={USA}}

 \affiliation[bbb]{organization={Department of Physics, Ben-Gurion University of the Negev},
             addressline={David Ben Gurion Boulevard 1},
             city={Beer Sheva},
             postcode={84105},
             country={Israel}}

\author[aaa]{Berndt Mueller}
\author[aaa]{Jyotirmoy Roy}
\author[bbb]{Sebastian Waeber}

\begin{abstract}
The description of heavy quarks inside the QGP medium, especially at low momenta, remains challenging from first principles, due to large coupling strength and gluon occupation numbers. An alternative way to formulate a description in terms of perturbative degrees of freedom is the interaction of the heavy quarks with the collective excitations of the medium in the form of phonons. This description is derived as an effective field theory (EFT) and therefore allows for a natural occurring power counting of the allowed interactions. This EFT has already been established in our previous work \cite{Kirchner:2025xdb}, where we have derived the phonon-heavy quark interactions and calculated their thermal scattering cross section.

In this work, we use these results, together with the heavy quark expansion, to calculate the heavy quark drag coefficient. Subsequently, we compare our results with the drag coefficient obtained from AdS/CFT calculations to determine the previously unknown Wilson coefficients for the quark-phonon coupling. Finally, we will use these coefficients to solve the Boltzmann equation for the heavy quark distribution, studying its thermalization in the QGP via soft phonon interactions.
\end{abstract}



\begin{keyword}
Heavy Quarks \sep Effective Field Theory \sep Thermalization
\end{keyword}

\end{frontmatter}



\section{Introduction}
\label{sec:Intro}

Heavy quarks with a mass much larger than the temperature of the QGP medium provide an excellent probe for the entire evolution of a heavy ion collision, since they are created in the first stages of a collision and therefore experience the full subsequent evolution including the expansion phase, freeze out and hadronic rescatterings. The description of the interactions between the heavy quarks and the medium is formulated in terms of QCD. For a large momentum transfer these interactions can be treated perturbatively. For lower momenta transfers between the heavy quark and the medium a description from first principles is more challenging due to high occupation numbers and a large coupling. To find a description amendable to this kinematic regime we will develop an effective field theory describing the interactions between the heavy quark and the effective degrees of freedom in the medium which will be its collective excitations.

We will begin with a short summary on Lagrangian fluid dynamics and then construct the EFT for the fluid-matter interactions. Next we will examine the quark-phonon interactions and then proceed to use them to study the energy loss and thermalization process of heavy quarks in the medium. A more detailed description of the results presented here can be found in \cite{Kirchner:2025xdb}.

\section{Construnction of the effective field theory}
\label{sec:Construction}

The description of fluid dynamics in terms of a field theory is based on the comoving coordinates of its fluid volume elements $\phi_I(t,\vec{x})$ with $I=1,2,3$ which track the movement of the individual fluid cells during its evolution. The ground state of the fluid at rest is then given by $\langle \phi_I(t,\vec{x}) \rangle = x^I$ (for all $t$). The guiding principle for the construction of the fluid EFT are the symmetries of the fluid, which are given by invariance under translations, rotations and volume preserving diffeomorphisms. The resulting minimal object invariant under these symmetries can then be constructed as
\begin{equation}
J^{\mu} = \frac{1}{3!}\varepsilon^{\mu \nu \rho \sigma} \varepsilon_{I J K} \partial_{\nu} \phi^{I} \partial_{\rho} \phi^{J} \partial_{\sigma} \phi^{K}.
\end{equation}
Finally, the Lagrangian of the fluid is given by an arbitrary function $F(X)$, encoding the microscopic dynamics of the fluid with $X=J^\mu J_\mu$. The function $F(X)$ can here be understood as equation of state, as it is linked to the thermodynamic variables such as the energy density $\epsilon=-F(X)$ which becomes evident when calculating the energy momentum of the theory \cite{Endlich:2010hf, Dubovsky:2011sj}.

Next we will introduce small fluctuations of the fluid around its equilibrium value $\phi^I = x^I + \lambda \pi^I(t,\vec{x})$. After a field redefinition $\pi^I \to \sqrt{-F'(1)} \pi^I$ we find the quadratic Lagrangian to be $\mathcal{L}^{(2)} = \frac{1}{2} (\partial_t \vec{\pi})^2 - \frac{1}{2} c_s^2 (\nabla \cdot \vec{\pi})^2$ with the speed of sound being given by $c_s^2  =(F'(1)+2F''(1))/F'(1)$. Evaluating the higher order terms of the Lagrangian, yields the $n$-phonon interaction vertecies which exist for any $n$. However, through the previous rescaling of the fields we are able to introduce the UV scale of the theory defined as $\Lambda_h = \sqrt[4]{-F'(1)}$, suppressing higher order interactions more strongly. To determine the precise value of the UV scale one would ideally carry out an analytic matching procedure to the UV theory. Since the UV theory in this case is given by non-perturbative QCD, this matching procedure is analytically not possible. Therefore, we will use kinematic considerations to get an estimate of the scale of $\Lambda_h$. A first constraint arises from the sound attenuation length in perturbative QCD and the strong coupling limit, which restricts the wave vector of the perturbation to be $k \lesssim 5T$ (pQCD) and $k \ll 11T$ (strong coupling) with $T$ being the temperature of the medium. Therefore, we will use $\Lambda_h \approx 5T$ in the following.

We can now construct the interactions between the fluid and the heavy quarks by considering all allowed terms, resulting in 
\begin{align}
		\mathcal{L} &= F(X) + Z_\psi(X)\Bar{\psi} i\slashed{D} \psi- M(X)\Bar{\psi}\psi + Z_A(X)G^a_{\mu\nu} G_a^{\mu\nu} +P(X) \Bar{\psi} \gamma_{\mu} \psi J^{\mu} + Q(X) \Bar{\psi} i \overleftrightarrow{D}_{\mu} \psi J^{\mu}  + R(X) \Bar{\psi} \sigma_{\mu\nu} \psi \omega^{\mu\nu}
\end{align}  
with parameters and coefficients such as the heavy quark mass now being promoted to functions of the dimensionless variable $X$ and $\psi$ and $A^\mu_a$ being the heavy quark and gluon fields, respectively\footnote{For more details on the terms appearing in the Lagrangian see \cite{Kirchner:2025xdb}}. Applying the phonon expansion to each term results in the interactions between phonons, heavy quarks and gluons, with the higher order interactions, such as a $n$-phonon heavy quark interaction being again suppressed by higher powers of the UV scale.

\section{Quark-Phonon Interactions}
\label{sec:Interactions}

\begin{figure}[t]
	\centering
	\includegraphics[width=0.7\textwidth]{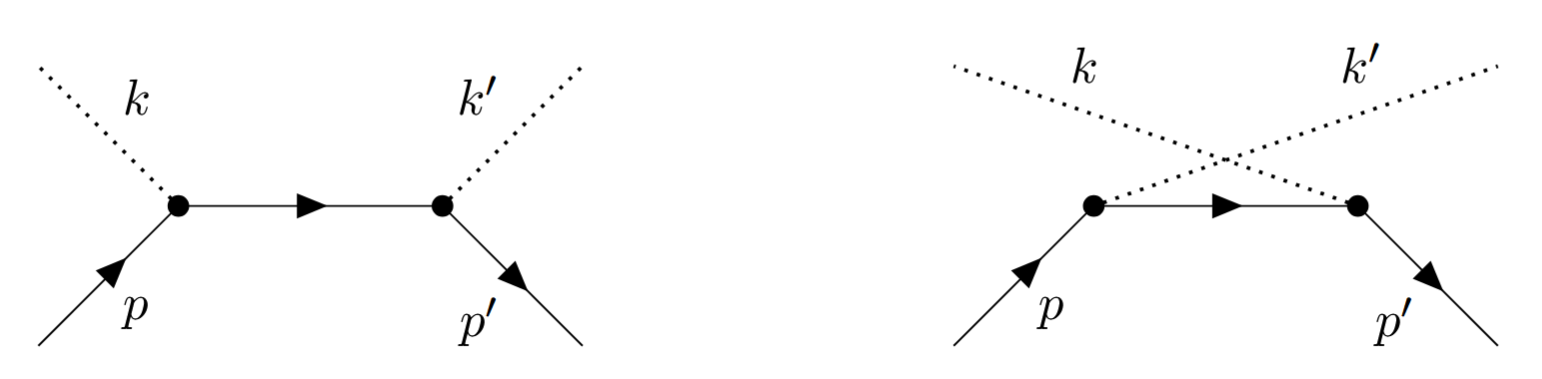}
	\caption{Leading order ($\Lambda_h^{-2}$) $2 \to 2 $ heavy quark-phonon scattering diagrams}\label{Fig:Feynman}
\end{figure}

\begin{figure}[h]
	\centering
	\includegraphics[width=7cm,clip]{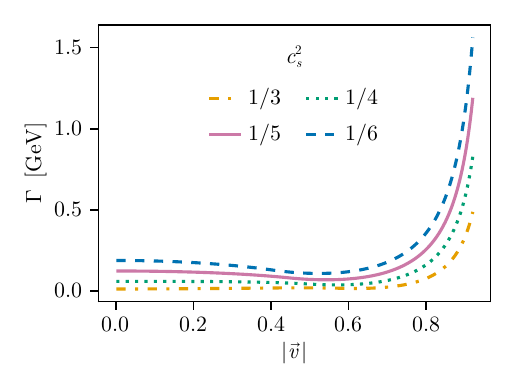}
	\includegraphics[width=7cm,clip]{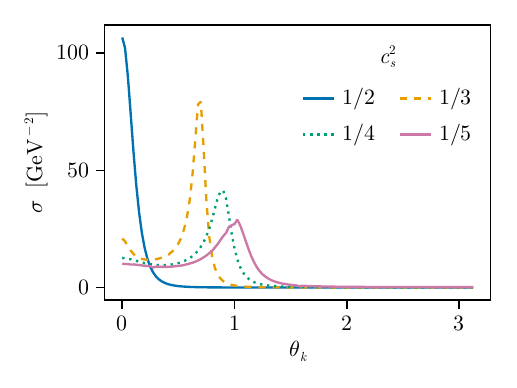}
	\caption{Quasi particle width as function of velocity for different values of the speed of sound (left) and heavy quark-phonon scattering cross section as function of the relative angle (right). The width is finite for all velocities and rapidly grows for a supersonic quark due to its Mach cone. The cross section is finite in the colinear limit and displays a resonance peak related to the Mach condition of the heavy quark moving through the medium.}\label{Fig:Width_XSec}
\end{figure}

We will now use the previously established interactions and examine the scattering between phonons and a heavy quark. The leading order $2\to2$ interaction is suppressed by $\Lambda_h^{-2}$ with its diagrams being shown in Fig.~\ref{Fig:Feynman}.
Since phonons are space-like particles, the heavy quark propagator can go on shell, requiring the use of a quasi particle width for the heavy quark. This width is obtained through the absorption of thermal phonons and the emission of phonons if the quark is moving above the speed of sound. The resulting quasi particle width is shown in Fig.~\ref{Fig:Width_XSec} (left panel).

We find that the decay width is non-zero at all velocities, ensuring a finite matrix element for the scattering process. However, when reaching large velocities of the heavy quark, the quasi particle width grows exponentially resulting in the effective field theory not being valid anymore due to the violation of unitarity. The sharp increase of the width at larger velocities is driven by the emission of phonons from the supersonic quark, i.e. by its Mach cone.


Using the calculated quasi particle width to regulate the divergence in the propagator of the matrix elements in Fig.~\ref{Fig:Feynman}, we find the heavy quark-phonon cross section depicted in Fig.~\ref{Fig:Width_XSec} (right panel).

\begin{figure}[t]
	\centering
	\includegraphics[width=7cm,clip]{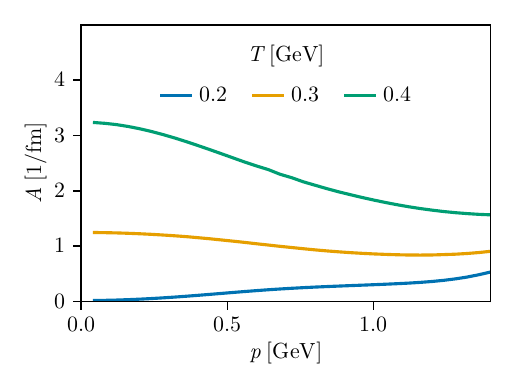}
	\includegraphics[width=7cm,clip]{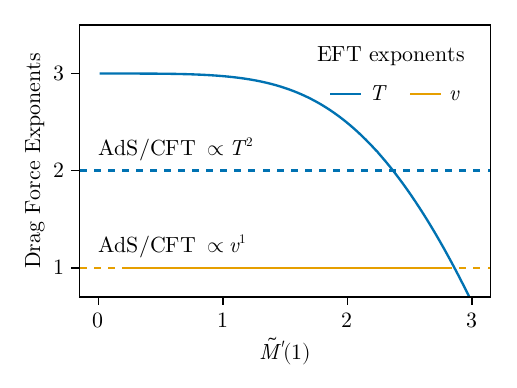}
	\caption{Heavy quark drag coefficient $A$ as function of the heavy quark momentum for different temperatures (left) and running of the temperature and velocity scaling of the drag force as function of the Wilson coefficient (right). }\label{Fig:A_Scaling}
\end{figure}

The heavy quark-phonon scattering cross section is finite in the co-linear limit due to the larger interaction time and vanishes for head-on collisions. The interaction displays a resonance peak with its position being linked to the speed of sound through the mach condition and scales inversely with the speed of sound. Note that this cross section depends on a so far undetermined Wilson coefficient taken to be one half here and in the following. Since this cross section is not observable in experiment we will use it to calculate quantities tied more closely to experimental observables in the following.

\section{Quark energy loss}
\label{sec:EnergyLoss}

We will begin by calculating the drag coefficient $A$ appearing in the Langevin equation 
\begin{equation}
	\dot{\vec{p}} = -A(\vec{p}) \vec{p} + \vec{\xi}
\end{equation}
with $\vec{p}$ being the momentum of the heavy quark and $\vec{\xi}$ being a random momentum kick from the medium \cite{Svetitsky:1987gq}. The resulting drag coefficient is depicted in Fig.~\ref{Fig:A_Scaling} (left panel).

While the absolute magnitude of the drag coefficient is not determined due to the undetermined Wilson coefficient for the heavy quark-phonon coupling, we already find that the heavy quark looses more energy in a hotter medium and at lower momenta. In order to determine this Wilson coefficient, we will use the drag coefficient to calculate the drag force on the heavy quark and match this to the low-momentum drag force calculated from AdS/CFT. Since the coupling in AdS/CFT is formally infinite, we will be using the scaling of the drag force as function of the temperature and velocity for the matching procedure and use $F_{\text{AdS/CFT}} \propto T^2v$. The resulting scaling of the EFT drag force as function of the Wilson coefficient is shown in Fig.~\ref{Fig:A_Scaling} (right panel). We find that the EFT drag force scales linearly with the velocity for all values of the Wilson coefficient, but shows a strong dependence for the temperature scaling. Using this scaling, we find the Wilson coefficient to be approximately $\tilde{M}'(1)=2.2$.

\begin{figure}[t]
	\centering
	\includegraphics[width=17cm,clip]{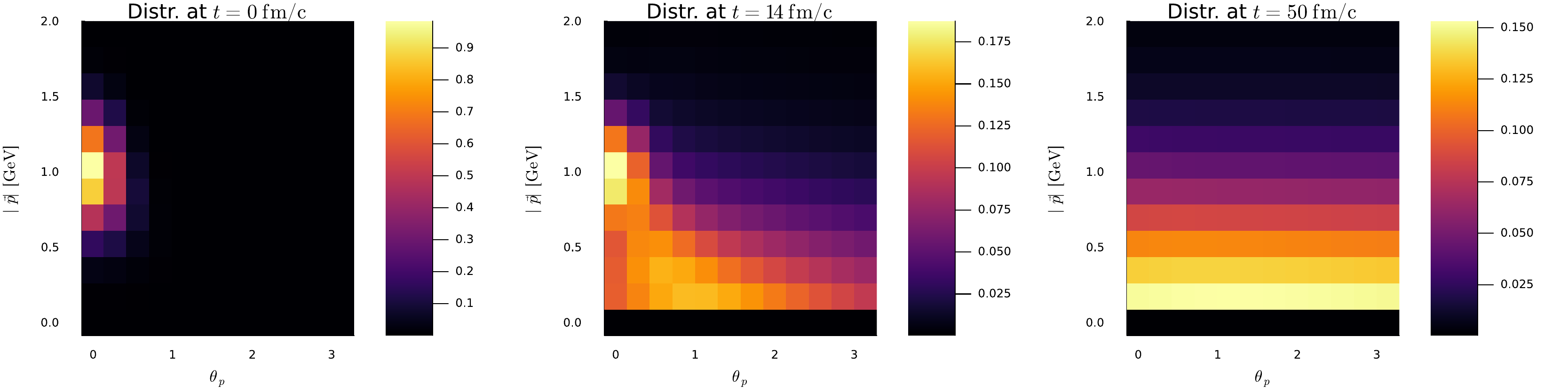}
	\caption{Evolution of the heavy quark distribution function for different time steps, starting from its initial condition (left panel) to the fully thermalized distribution (right panel) as function of the heavy quark azimuthal momentum angle and momentum magnitude.}\label{Fig:Boltzmann}
\end{figure}

Finally, we will use this Wilson coefficient to study the thermalization process of the heavy quark using the Boltzmann equation. We will assume a spatially homogeneous medium and a spatially homogeneous distribution of the heavy quarks allowing us to write the Boltzmann equation as
\begin{equation}
	\partial_t f(t,\vec{p}) = C[f](t,\vec{p})
\end{equation}
with $f(t,\vec{p})$ being the heavy quark distribution function and $C[f]$ being the collision kernel calculated from the $2\to2$ scattering matrix element Fig.~\ref{Fig:Feynman}. We initialize the heavy quark distribution with a Gaussian in momentum space centered around $\vec{p}_0 =(0,0,1 \; \mathrm{GeV})$, i.e. with the quarks all moving along the $z$-direction (left panel, Fig.~\ref{Fig:Boltzmann}). The distribution then becomes more isotropic through interactions with the medium (central panel, Fig.~\ref{Fig:Boltzmann}) and is fully isotropic and thermalized in the long time limit (right panel, Fig.~\ref{Fig:Boltzmann}).

\section{Conclusion}
\label{sec:Conclusion}

In this work we have developed an effective field theory describing the heavy quark dynamics in the medium via interactions with phonons, for which we find a naturally appearing power counting. We have used this theory to derive its Feynman rules, which in turn were used to calculate the quasi particle width of the heavy quark and its scattering cross section with phonons. Using this scattering cross section, we were able to calculate the drag on the heavy quark and determine the coupling to the medium via comparison with AdS/CFT. Finally, we were able to study the evolution of the heavy quark distribution solving the Boltzmann equation, demonstrating its thermalization through interactions with the medium. In future work, we want to extend the theory through the inclusion of viscosities and use this framework to study the gluon momentum broadening.

\bibliographystyle{elsarticle-num}



\end{document}